\documentclass[10pt,twocolumn]{article}
\usepackage[margin=0.65in]{geometry}
\usepackage[T1]{fontenc}
\usepackage[utf8]{inputenc}
\usepackage{lmodern}
\usepackage[english]{babel}
\usepackage{graphicx}
\usepackage{amssymb}
\usepackage{overpic}
\usepackage{pdfpages}

\usepackage{algpseudocode}
\usepackage{algorithm}

\usepackage[unicode]{hyperref}
\hypersetup{
    colorlinks=true,
    linkcolor=black,
    citecolor=black,
    filecolor=black,
    urlcolor=black,
    linkbordercolor = {1 1 1},
}

\usepackage[square,numbers]{natbib}

\title{Estimating savings in parking demand using shared vehicles for home-work commuting\footnote{Published in IEEE Transactions on Intelligent Transportation Systems.\newline \copyright 2018 IEEE. Personal use of this material is permitted. Permission from IEEE must be obtained for all other uses, in any current or future media, including reprinting/republishing this material for advertising or promotional purposes, creating new collective works, for resale or redistribution to servers or lists, or reuse of any copyrighted component of this work in other works. \newline DOI: \href{https://doi.org/10.1109/TITS.2018.2869085}{10.1109/TITS.2018.2869085} }}
\author{D\'aniel Kondor$^{1,2}$, Hongmou Zhang (
\makebox{\raisebox{-1.5pt}{\includegraphics{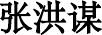}}}%
)$^{1,3}$, Remi Tachet$^1$, Paolo Santi$^{1,4}$, Carlo Ratti$^1$\\
		\normalsize{$^1$Senseable City Laboratory, MIT, Cambridge MA 02139 USA}\\
		\normalsize{$^2$Singapore-MIT Alliance for Research and Technology, Singapore}\\
		\normalsize{$^3$Department of Urban Studies and Planning, MIT, Cambridge MA 02139 USA}\\
		\normalsize{$^4$ Istituto di Informatica e Telematica del CNR, Pisa, Italy}\\
		\normalsize{E-mail: \texttt{dkondor@mit.edu}}}
\date{\today}
	
\begin{document}

	\maketitle
	
	\begin{abstract}
		The increasing availability and adoption of shared vehicles as an alternative to personally-owned cars presents ample opportunities for achieving more efficient transportation in cities. With private cars spending on the average over 95\% of the time parked, one of the possible benefits of shared mobility is the reduced need for parking space. While widely discussed, a systematic quantification of these benefits as a function of mobility demand and sharing models is still mostly lacking in the literature. As a first step in this direction, this paper focuses on a type of private mobility which, although specific, is a major contributor to traffic congestion and parking needs, namely, home-work commuting. We develop a data-driven methodology for estimating commuter parking needs in different shared mobility models, including a model where self-driving vehicles are used to partially compensate flow imbalance typical of commuting, and further reduce parking infrastructure at the expense of increased traveled kilometers. We consider the city of Singapore as a case study, and produce very encouraging results showing that the gradual transition to shared mobility models will bring tangible reductions in parking infrastructure. In the future-looking, self-driving vehicle scenario, our analysis suggests that up to 50\% reduction in parking needs can be achieved at the expense of increasing total traveled kilometers of less than 2\%.
	\end{abstract}

	
	\section{Introduction}
		
		{T}{raffic} caused by privately owned vehicles presents major challenges in urban environments around the world, with 
		pollution and congestion being serious concerns. Part of the problem of congestion is the high amount of space cities need to dedicate 
		to roads, parking lots and garages, posing problems in high-density downtown areas and having a huge impact on shaping suburban 
		communities, where planning is often centered around cars and parking spaces. As an example,~in car-dependent Los Angeles county, roads take up 
		about 140 square miles, while parking spaces in total take up 200 square miles; this latter area is equivalent to about 14\% of all 
		incorporated area in the county~\cite{Chester2015}. Apart from a variety of regulatory and development policies governments use in
		response to challenges associated with urban transportation~\cite{Flyvbjerg2007,Mercier}, it has been shown that specific policies
		on parking have substantial effects on urban areas~\cite{Shoup,Weinberger2012,McCahill2016,Fiez2017}.
		
		After rapid technological developments especially over the past decade, autonomous vehicle (i.e.,~\emph{self-driving}) technology is
		expected to be ready for wide deployment in the near future with large implications for urban mobility~\cite{Fagnant2015,Krueger2016,
		Harper2016,Daziano2017}. It is generally accepted that one of the main benefits of self-driving cars could be reduced road congestion,
		as current roads are expected to have much higher capacity if the majority of traffic is autonomous vehicles~\cite{VandenBerg2016}. On
		the other hand, the convenience of autonomous vehicles can generate significant further traffic, both from people who currently are not
		able or prefer not to drive, and more generally as well, similarly to how increasing road and parking capacity often leads to increased 
		traffic~\cite{Harper2016,Handy2005,Gupta2006,Smith2012}.
		
		Further gains are expected from using shared autonomous vehicles instead of private ones, with people buying \emph{mobility-as-a-service}
		instead of cars~\cite{Firnkorn2012}. A major expected benefit of a shared car system is better economics: the cost of owning and
		maintaining vehicle can be distributed proportionally among the per-trip costs, allowing people to make more informed choices about
		their transportation mode and vehicle type on a much more granular level. Furthermore, as private cars are parked most of the time, it
		is expected that a smaller fleet of better utilized shared vehicles could service the same mobility demand, offering reductions in need
		for parking as well~\cite{Fagnant2014,Fagnant2015,Burns2013}. On the other hand, since per-trip costs are expected to be significantly
		lower than current costs of trips made with either a private car or with a taxi service, the availability of a fleet of shared autonomous
		vehicles can again lead to significant increase in total traffic volume as people currently not being able to afford a car can switch
		from alternative modes of transportation~\cite{Burns2013,Brownell2014,Smith2012}.
		
		Apart from the expectations from autonomous vehicles, car-sharing has been proposed as a more efficient alternative to private car
		ownership decades ago, while large-scale deployment only occurred in the past 15-20 years, mainly due to advances in smart technologies~%
		\cite{Jorge2013}. Proponents argue that many benefits of sharing can be achieved with conventional shared cars, while there are practical
		challenges limiting adoption, including user anxiety about finding a nearby vehicle or parking and the problem of rebalancing if one-way
		trips are allowed~\cite{Kek2006,Papanikolaou2011}. These problems could be easily solved with autonomous vehicles; thus we expect that
		in the future the distinction between taxi, ride-sharing, car-sharing services and even transit will blur and new, integrated solutions will
		become possible, providing services similar to personal rapid transit systems proposed but never implemented in the previous century~%
		\cite{Anderson2000,Brownell2014}. Consequently, we find it important to characterize the expected performance of transportation
		solutions based on the sharing of vehicles.

		\subsection{Contributions}
		In this paper, we focus on commuting between home and work, and investigate the possible gains from car-sharing and self-driving on
		the number of parking spots and vehicles required. Contrary to previous studies, we focus specifically on commuters who contribute a
		major portion of road traffic and parking demand, yet are not the typical target of car-sharing or even taxi services. A reason for
		this is that commuting flows are typically imbalanced and traffic demand is highly concentrated in rush hours. These factors make
		regular commuters a difficult target for current commercial car-sharing solutions, ride-sharing and taxi services; on the other hand,
		due to the large amount of traffic associated with commuting, even moderate gains in efficiency can have large benefits for cities.
		Additionally, as there are well established methods to estimate commuting flows from mobile phone usage data, our methodology can be
		easily applied to provide baseline estimates of possible efficiency gains, in contrast to more detailed case studies which would require
		accurate data on general purpose trips. We specifically focus on parking, as the decrease in parking needs is expected to be a clear
		positive outcome; we note that an estimate of decrease in the total number of cars is less meaningful, since each vehicle is expected
		to travel more, potentially giving rise to similar levels of congestion as private cars today. On the other hand, focusing on parking
		captures a potential benefit from smaller total fleet sizes.
		
		We use data from mobile phone network logs to estimate home and work locations for a large sample of the population in Singapore and
		simulate their daily trips assuming private, shared and shared self-driving car usage. In the case of shared cars driven by their users,
		a main limiting factor for sharing is that the car needs to be parked at a comfortable walking distance from the origin and destination
		of their users. In the case of self-driving, this limitation is removed as the car can be allowed to travel longer distances to a
		parking spot or their next customer, at the expense of higher total vehicle miles traveled (VMT); we explore the implications of this
		trade-off by varying the distance self-driving cars are allowed to travel without a passenger. Furthermore, we repeat simulations with
		varying presumed adoption rates to estimate which rate is required to gain sizable benefits.
		
		We note that a main limitation in our approach is that, beside taking note of any additional distance traveled, we do not explicitly
		model any effect on congestion as that would require a detailed microsimulation of traffic and assumptions on the actual performance of
		autonomous vehicles in real traffic conditions. Furthermore, while we expect that people's behavior will change in response to
		availability of shared and self-driving vehicles, we do not aim to model this in our current work yet; we only assume that a certain
		share of commuting is made with shared vehicles.
		
		Summarizing, the novel contribution of this paper is the development of a methodology that, starting from extensive real-world mobility traces, provides an accurate estimation of parking needs in a variety of sharing scenarios, including the effect of self-driving vehicles.
		
		\subsection{Related work}
		In accordance with the growing adoption of car sharing and the potential impact of self-driving, there is a significant research interest
		in assessing potential effects with regards to usage patterns, traffic and emissions. Survey-based methods find that car ownership among
		car-sharing users decreases significantly, up to 40\%, depending on the study and the parameters used to correct for sampling effects~%
		\cite{Martin2010,Firnkorn2012}. Still, drawing conclusions for the more wide-spread adoption of shared vehicles is not straightforward,
		since current car-sharing users are probably not a representative sample of the general population. Further studies try to estimate
		public attitude toward mobility options represented by self-driving vehicles and estimate the potential for adoption based on these~%
		\cite{Krueger2016,Harper2016,Daziano2017}. Several studies then try to estimate the fleet size which could serve a certain population given some
		operational parameters, and the associated costs for travelers. Studies based on randomly generated trips find that about 10\% -- 15\%
		of cars could serve mobility demands compared to private vehicles, with significantly reduced costs when compared to either privately
		owned cars or taxi rides~\cite{Fagnant2014,Burns2013,Brownell2014}. This also prompted some concerns about the possibility of shared
		self-driving cars inducing significantly more traffic since they offer much cheaper and more convenient means of transportation~\cite{
		Smith2012,Harper2016}. A more recent study based on realistic origin-destination flows obtained from travel surveys in Singapore and a 
		theoretical derivation for fleet size finds that a fleet which has a size of about 38\% of the number of privately owned vehicles can
		satisfy mobility demand with a bound of 15 minutes on passenger waiting times~\cite{Ballantyne2014}. Further work in the central area
		of Singapore focused on the trade-off between fleet size and utilization using a detailed simulation of people's mobility~\cite{
		Marczuk2015}.
		
		Concentrating on parking, a recent study has shown that by utilizing space much more efficiently, AVs have the potential to 
		significantly reduce to spatial footprint of parking facilities~\cite{Nourinejad2018}. Regarding shared vehicles, our work is most 
		similar to studies by Zhang~et~al.~\cite{Zhang2015,Zhang2017}, who find that parking demand could be reduced by up to~90\% for people 
		switching to shared autonomous vehicle usage. The main difference is that the authors in~\cite{Zhang2017} focus on the use of 
		\emph{existing} parking infrastructure, while in the current study we aim to calculate minimum parking requirements based only on basic 
		assumption about commuter behavior, thus our methodology does not require any previous knowledge of available parking which can be 
		difficult to obtain, especially on large scales~\cite{Zhang2017,Chester2015}. Furthermore, our simulation includes a significantly 
		larger target population and more than 100~times larger fleet size (while the authors in~\cite{ Zhang2017} only consider 5\% of the 
		population of the city of Atlanta, i.e. about 22~thousand people in total, we consider a sample of over 1 million commuters in 
		Singapore). A further recent study investigating the operational characteristics of a shared autonomous vehicle system in Lisbon, 
		Portugal also considered potential reductions in parking needs with estimating that all on-street parking and a significant amount of 
		off-street parking could be eliminated~\cite{OECD2015}.

	\section{Methods}

	\subsection{Home and work location detection}
		
		For the purpose of this work, we use call record detail records (CDRs) provided by Singtel, the largest mobile network operator in
		Singapore. The data includes records of several million subscribers for a period of eight weeks. The data includes a record when a user
		places or receives a call, or sends or receives a text message; data connections or handover information is not included. Each record
		includes the location of the antenna handling the event; with the high density of antennas in Singapore, spatial accuracy is estimated to
		be around a few hundred meters. Our dataset does not allow the reconstruction of individual trip data, but can be efficiently used to
		detect home and work locations of mobile phone users; this is considered standard and well-established practice~\cite{Bagrow2012,Jiang,%
		Alexander2015}. 
		
\begin{figure}
			\centering
			\includegraphics[width=3.2in]{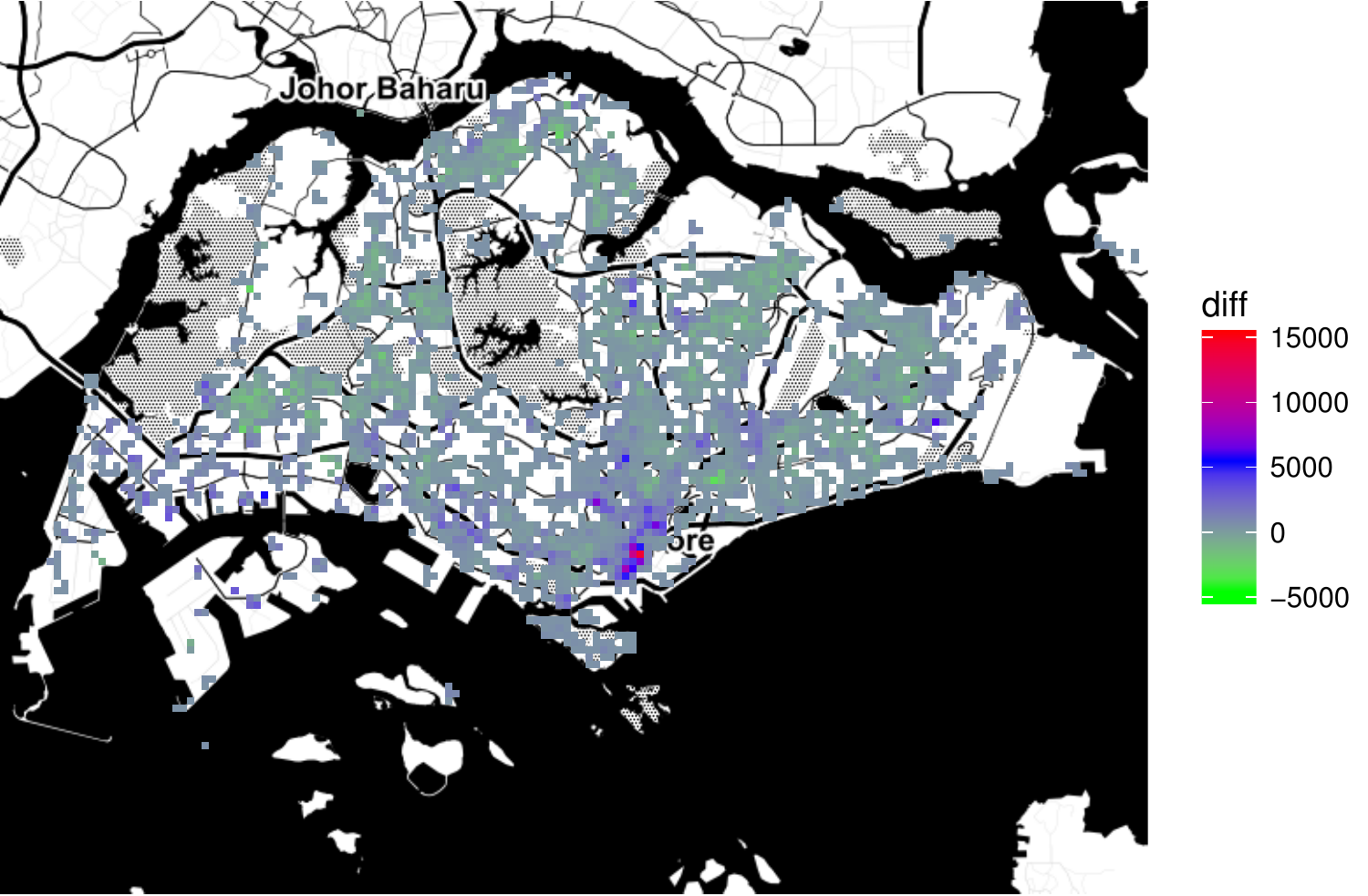}
			\caption{Distribution of difference of number of work and home locations (red means more work locations, while green means more home
				locations); these differences set a limit on the minimum needed parking spaces.}
			\label{diffdist}
		\end{figure}

		Clustering people's locations and identifying the main nighttime and daytime clusters result in our estimates on home and work locations.
		To ensure the quality of the results, we use the criteria that the clusters identified as work or home locations should have at least 20
		records during working hours or during evenings and at night respectively. Furthermore, for the following work, we only include people whose
		identified home and work locations are at least 1~km distance apart (using simple geodesic distance) and thus are possible candidates for
		commuting by car. There are a total of 1,992,950 people in the dataset whose home and work locations could be reliably detected, and
		1,066,504 of these fulfill the criteria that the two locations are more than 1~km apart. We show the obtained spatial distribution of home
		and work locations in Fig.~S1 and the distribution of commute distances in Fig.~S2 in the Supplementary Material. Furthermore, we display
		the difference between home and work locations in Fig.~\ref{diffdist}; as unbalanced flows in the morning and evening present a fundamental
		challenge to sharing cars and parking spaces, this will pose an inherent limit to the possible gains in efficiency from them. Since the
		granularity of detected locations is that of antennas in the network (i.e.~each location corresponds to an antenna), we add a random noise
		of the magnitude of $166\,\mathrm{m}$ to users' locations so that these will be less clustered. We note that the main assumption behind
		the current work is that the home and work locations obtained from this dataset will be a representative sample of people who would choose
		commuting by car.
		
	\subsection{Travel times}
		
		In order to better estimate commute times, we calculate the travel times between people's home and work locations based on 
		real-world data as well. In the case of Singapore, average travel times between a set of road intersections were provided by the Land 
		Transport Authority, measured at different times of the day and week. There are a total of 11,789 intersections, providing a good 
		coverage of the area. For each user in the dataset, we located the closest intersection to their home and work location and use the 
		travel time between these points as an estimate. We use estimates for times between 7AM and 8AM in the morning for travel from home to 
		work and estimates for times between 4PM and 5PM as for travel from work to home. We display the distribution of these (as compiled for 
		the list of people in the dataset) in Fig.~S3. The travel time distributions have a mean of $1199\,\mathrm{s}$ and $1027\,\mathrm{s}$ 
		respectively for the morning and afternoon case, while the medians are $1090\,\mathrm{s}$ and $983\,\mathrm{s}$. Note that these seem 
		relatively low when comparing to typical values people spend by daily commuting. We speculate that this is the effect of Singapore's 
		highly restrictive policy on private car ownership, but highly car-centric road infrastructure, resulting in cars being a highly 
		efficient means of transport for those who can afford them\footnote{In 2010, there were about 780 thousand private cars in Singapore, a 
		city with a population of about 5 million (3.2 million citizens and 1.8 million permanent residents and visitors), giving a ratio of 
		only 154 cars per 1000 population (241 per 1000 when only counting citizens); this is significantly lower than the value of 500 -- 800 
		found in other developed countries. This is mainly achieved by the government setting quotas on newly registered vehicles and auctioning 
		spots to potential buyers. In October 2017, as the result of the auctioning, the levy to register a new car for a 10-year period was 
		about S\$41,000 (US\$31,000).}.

	\subsection{Simulated scenarios}
		
		In this work, we focus on a set of commuters as described in the previous section and estimate the number of parking spaces and vehicles
		needed to satisfy their mobility demand. In the following, we denote the number of users in our dataset by $N_U$, the number of required
		parking spaces by $N_P$, and the number of required cars by $N_C$. Furthermore, we measure the total distance traveled by commuters, denoted
		by $d_\mathrm{tot}$. We employ several scenarios for their commuting habits and compare the results and quantify the improvement due to
		sharing vehicles and self-driving:
		
			\paragraph*{1) No sharing} Each person uses a private car and has a private reserved parking space at their home and work location. In
				this case, it trivially follows that $N_C = N_U$ and $N_P = 2 N_C$, while $d_\mathrm{tot}$ is simple the sum of distances between
				people's home and work locations.
			\paragraph*{2) Private cars, shared parking} In this scenario, all parking is shared, with people taking the closest available spot to
				their destination at the end of each trip. Furthermore, we require that everyone is able to find parking closer than a given
				$r_\textrm{max}$ to their destination, which is the main parameter in the simulation. In this case, $N_C = N_U$,
				$N_C \leq N_P \leq 2 N_C$, while the total distance traveled ($d_\mathrm{tot}$) will increase as people have to
				reach their actual parking spot from their destination.
			\paragraph*{3) Shared vehicles} In this case, we assume that everyone is using shared cars to commute to work. This means that people
				always take the closest available car at the origin of their trip and park it at the closest available spot at the destination
				of their trip, with the requirement that cars and parking have to be available closer than $r_\textrm{max}$ to the origin and
				destination of each trip, respectively. The main gain in this case is that one vehicle can potentially complete more than two
				trips per day, thus $N_C \leq N_U$, while we still have $N_C \leq N_P \leq 2 N_C$.
			\paragraph*{4) Shared self-driving vehicles} In this case, it is assumed that the shared cars are capable of self-driving, thus they
				can pick up and drop off passengers at their exact home and work locations and then find an available parking spot in the
				neighborhood. Computationally, this case can be modeled in exactly the same way as the previous one, but an important difference
				is that $r_\textrm{max}$ now represents the distance self-driving cars are allowed to travel without a passenger. Thus, much
				larger values for $r_\textrm{max}$ are possible with the trade-off of adding extra traffic and further increasing $d_\mathrm{tot}$.
		
		We note that currently most cities have a mix of scenarios \#1 and \#2. Curb parking typically contributes to \#2, while most larger
		employers who provide on-site parking contribute to \#1, i.e.,~their garages are not utilized in any manner beside employee parking.
		Furthermore, many car owners prefer to have their designated spot at home if they can afford it (either a private garage, driveway or a
		reserved space in a parking lot or garage), which is then left underused during the day, but guarantees convenient parking when they
		arrive home in the evening. While our current work only assumes commuting between work and home, and thus the number of parking spaces
		per car is maximum two, in real cities the number of total available parking per car can be as high as $3.3$~\cite{Chester2015}.
		
		In contrast to scenario~\#1, the use of shared parking with conventional vehicles (scenarios~\#2 and~\#3) present additional anxiety to
		users about finding parking close to their destination to avoid excess walking. On the other hand, scenario~\#4 presents the convenience
		of picking up and dropping off passengers at their exact preferred locations, which can be a substantial advantage over both private or
		shared conventional vehicles.
		
	\subsection{Computational implementation}
		
		\begin{algorithm}[t]
		\begin{algorithmic}
			\State $E$ = \{ list of trip events; one event is either the start or end of a trip \}
			\State $r_\textrm{max}$ = maximum distance people are willing to walk
			\State $N_P$ parking spaces required (initially one per person)
			\State $L_P$ = \{ list for free parking spaces \}
				\State process all events in $E$ in time order:
				\ForAll{$e \in E$}
					\If{$e$ is the start of a trip}
						\State add to $L_P$ a new empty parking space
						\State \, with $e$'s the coordinates
					\Else \, $e$ is the end of a trip
						\State find closest free parking space $p \in L_P$
						\State \, s.t. $dist(e,p) < r_{\textrm{max}}$
						\If{found}
							\State remove $p$ from $L_P$
							\State (i.e.~user occupies $p$)
							\State start the user's next trip from $p$
						\Else
							\State assume there is a more parking
							\State increase $N_P$ by one
							\State start the user's next trip from $e$
						\EndIf
					\EndIf
				\EndFor
			\State Result: $N_P$ total number of parking spaces needed to satisfy mobility demand
		\end{algorithmic}
		\caption{\small Main algorithm to calculate parking demand for private vehicles with shared parking for a set of trips generated for one
			day (scenario \#2 above). The event list $E$ is generated with assigning random start timestamps to every person's home-work and
			work-home trips.}
		\label{estimate_private}
		\end{algorithm}
		
		We run simulations to determine the demand for parking spaces and the opportunities for sharing in scenarios \#2 -- \#4 and compare
		results to the constant values in the case of scenario \#1. We show the simulation algorithm in the case of private vehicles (\#2) as
		Algorithm~\ref{estimate_private} and for shared or self-driving vehicles (\#3 or \#4) as Algorithm~\ref{estimate_shared}. In both cases,
		the input is a set of trips (generated from the home and work locations) and potentially a set of free parking spaces and available
		shared vehicles (only for Algorithm~\ref{estimate_shared}).
		
		In the case of private vehicles (\#2) in Algorithm~\ref{estimate_private}, we start the simulation with assuming that everyone has a
		parking spot at their home location and do not assume any more parking spaces at work locations yet. In accordance with this, we set
		the total number of parking spots in the city to be $N_P = N_C = N_U$, and the set of available parking ($L_P$) is empty. At first, as
		people leave home in the morning, their home parking spots become available for other to use. We keep track of free parking spots in the
		list $L_P$ (employing a spatial index for efficient searches later). When someone arrives at their work location, they search for free
		parking spots in $L_P$ within a $r_\textrm{max}$ radius. If such a parking spot is found (i.e.,~someone's home spot that is unused), it
		can be occupied; in case of more than one parking available within the search radius, we always select the closest one. If there are no
		free parking spots close to an arriving person's work location, we add one more which they occupy. Thus, we increase the number $N_P$ of
		parking spots by one. We can assume that this parking spot was
		there all the time, but no one needed it yet. When moving people back home, we repeat the same procedure: everyone takes their car from
		where they parked it in the morning (adding that spot to $L_P$), drives home and tries to find a free spot. Since leaving from work and
		arriving at home happens stochastically, it can happen that a person finds their ``home'' spot occupied. In this case, they again search
		for the closest available alternative spot, or if none is found within an $r_\textrm{max}$ radius, we again add a further parking space
		to the city, again increasing $N_P$. Depending on the timing of commutes, people leaving and arriving at either their home or work
		locations happens interleaved, meaning that not all parking becomes available. This way, the timing of trips plays a significant role in
		the result as well. See the next subsection for assigning trip timings.
		
		In the case of car-sharing (\#3) and self-driving vehicles (\#4), as displayed in Algorithm~\ref{estimate_shared}, we not only maintain
		a list of free parking spots ($L_P$), but also of available vehicles, again including the coordinates where they are parked ($L_C$).
		When someone starts a trip, we first search in the list of available cars ($L_C$), and if a suitable car $c$ is found within $r_\textrm{max}$
		distance of the origin of the trip, we select the closest such car $c$, remove it from $L_C$ and add its location to $L_P$ as a free parking
		spot. On the other hand, if no such cars are found, we add one more car to the system at the trip origin location, increasing the total
		number of cars $N_C$. We also
		increase the number of parking spaces $N_P$ as we assume the newly added car to have been parked in that location, which again becomes a
		free parking spot and is added to $L_P$. In this case, at the beginning of the simulation, we do not place any parking spaces or cars in
		the system, i.e.,~we start with $N_P = N_C = 0$ and the $L_P$ and $L_C$ lists being empty. This way, during the course of the simulation,
		only the necessary number of vehicles and parking spaces are added. In this case, we also take into account the extra trip time due to
		traveling between the origin or destination of a trip and the parking location. This quantity can become significant for self-driving
		vehicles, especially if we consider a relatively larger $r_\textrm{max}$ radius.
		
		In all cases, it is assumed that the agents are able to find the closest available parking and closest available car when using shared
		cars. While searching for parking is complex problem by itself~\cite{Shoup2006,Dowling2017}, our assumption basically means that all
		drivers use an efficient navigation system which also receives real-time updates on parking availability. Implementing such a system
		is possible already with today's technology; also, we expect that shared autonomous vehicles will be able to communicate with a
		``controller'' that directs them to the closest available parking.
		
		Since the actual timing of morning and afternoon trips can affect the results -- see subsection below --, in case of both algorithms we
		run the simulation for multiple days in a row with different, randomly generated trip timings each day. It is important to note that we
		start with empty $L_P$ and $L_C$ lists only on the first day of the simulations; on subsequent days,
		we start the simulation with the $L_P$ and $L_C$ lists and $N_P$ and $N_C$ values obtained by the end of the previous day. This way, we
		are testing if the same number of parking, vehicles and actual spatial configuration of parking is sufficient to satisfy the travel
		demand on the next day, or if further parking and cars need to be added to the system to account for a different sequence of trips. In
		the experiments reported below, we ran the simulation for $n_d = 30$ days in each case.
		
		\begin{algorithm}[bh!]
		\begin{algorithmic}
			\State $E$ = \{ list of trip events; one event is either the start or end of a trip \}
			\State $r_\textrm{max}$ = maximum distance that
			\State \quad people are willing to walk (\#3 case) or
			\State \quad self-driving cars travel empty (\#4 case)
			\State $N_P = 0$ parking spaces required
			\State $N_C = 0$ number of cars required
			\State $L_P$ = \{ list for free parking spaces \}
			\State $L_C$ = \{ list for available cars \}
				\State process all events in $E$ in time order:
				\ForAll{$e \in E$}
					\If{$e$ is the start of a trip}
						\State find closest $c \in L_C$ s.t. $dist(e,c) < r_\textrm{max}$
						\If{found}
							\State remove $c$ from $L_C$
							\State add $c$'s location to $L_P$
							\State add travel time between $c$
							\State \, and $e$ to the total trip time
						\Else
							\State assume there is a free car at $e$
							\State increase both $N_P$ and $N_C$ by one
							\State add $e$'s location to $L_P$
						\EndIf
					\Else \, $e$ is the end of a trip
						\State find closest $p \in L_P$ s.t. $dist(e,p) < r_\textrm{max}$
						\If{found}
							\State remove $p$ from $L_P$
							\State add travel time between $e$
							\State \, and $p$ to the total trip time
							\State add $p$'s location to $L_C$
						\Else
							\State assume there is a more parking
							\State increase $N_P$ by one
							\State add $e$'s location to $L_C$
						\EndIf
					\EndIf
				\EndFor
			\State Result: $N_P$ total number of parking spaces and $N_C$ total number of cars needed to satisfy mobility demand
		\end{algorithmic}
		\caption{\small Main algorithm to calculate parking demand for shared or self-driving vehicles with shared parking (scenarios \#3 and \#4 above)
			for one day. Again, the event list $E$ is generated with assigning random start timestamps to every person's home-work and work-home
			trips.}
		\label{estimate_shared}
		\end{algorithm}
		
	\subsection{Simulation parameters}
	
		For all scenarios \#2--\#4, the main parameter that will affect the results is the bound $r_\textrm{max}$ on the distance between trip
		origin and destination and the sought parking spot. In the case of scenarios~\#2 and~\#3, this bounds represents the distance people are
		willing to walk from their parking location and their destination. In case of shared vehicles (scenarios~\#3 and \#4), $r_\textrm{max}$
		is also the upper bound to the distance to the closest available shared car. For scenario~\#4, this is the distance that self-driving
		cars are allowed to travel without a passenger before the start or after the end of a trip to reach their parking location.
		
		The second main parameter in the simulation is the method used to generate commute timings. This is represented by a \emph{commute window}
		of length $t_W$; all trips are assumed to start inside this window (see below). Furthermore, results are affected by the penetration ratio
		of shared mobility, i.e.,~the number of people who use shared or self-driving cars among the group of commuters considered.

	\subsection{Generating trip starting times}
		
		As we commented above, a main determinant on the possible efficiency gains is the sequence and timing of individual trips, since it
		determines if a specific shared vehicle or parking spot is available at the time when a commuter would want to start or finish their
		journey. Since timings of individual trips on a large scale are hard to obtain, and are still subject to daily variations, we generate
		these randomly for each person in the simulation. To test for variations in different realizations, we run the simulations for $n_d = 30$
		consequtive days and then repeat the whole process 100 times for better stochastic accuracy. Each day in a single simulation run presents
		a different realization of random trip start times. Running the simulation for several days helps establish the robustness of spatial
		configuration of parking and vehicles, while repeating the simulation allows us to test for statistical variations. We find that random
		variations are very small: standard deviation are less than $1\%$ in all cases, and less than $0.1\%$ in most cases. We report the effect
		of these variations in the Supplementary Material, in Figs.~S5, S7, S8 and~S9 and in Table~S1\footnote{Available as a separate download
		\url{https://www.dropbox.com/s/s410s74oxh28h07/parkefficiency_si_table_S1.ods?dl=0}}.
		
		For the main results of the current work,
		we generate the start time of each individual trip uniformly at random in a time window of length $t_W = 1\,\mathrm{hour}$, from 7AM to
		8AM for morning commutes and between 4PM and 5PM for afternoon commutes. Beside the main results, we further explore several options for
		$t_W$ and also an option where we generate trip start times based on a dataset of public transportation usage in Singapore. 
		
	\section{Results}
	
	\subsection{Reduction in parking spaces and cars required}
	
		\begin{figure*}
			\centering
			\includegraphics{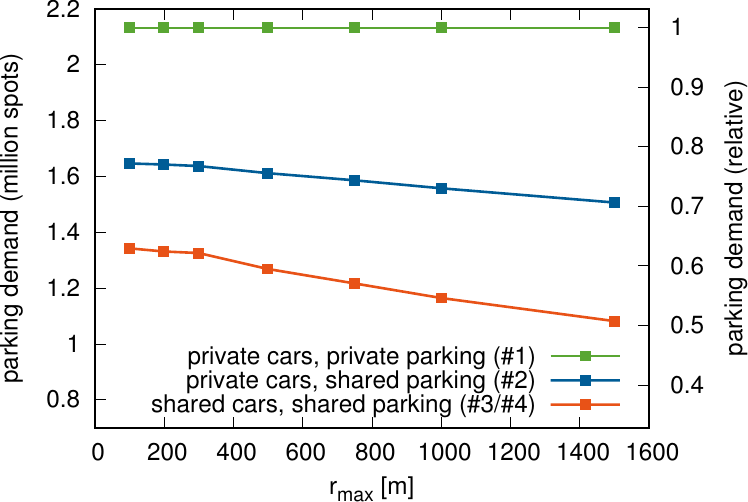} \quad \quad \quad
			\begin{overpic}{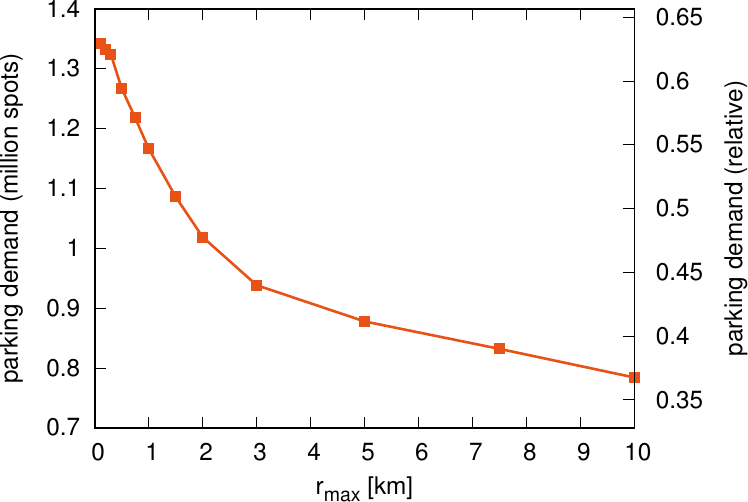}
				\put(34,29){\includegraphics{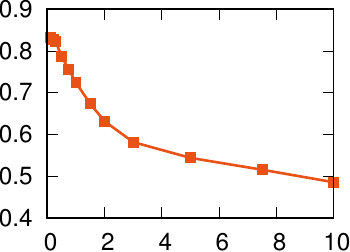}}
			\end{overpic}
			\caption{Comparing demand for parking in different scenarios. Left: Parking demand in scenarios \#1-\#4 up to
				$r_\mathrm{max} = 1.5\,\mathrm{km}$. The number of parking spaces required is displayed on the left $y$-axis, while relative
				numbers (compared to case \#1, i.e.~private parking spaces) are shown on the right axis. Right: parking demand in scenario
				\#4 with $r_\mathrm{max}$ values up to $10\,\mathrm{km}$. The inset shows relative numbers compared to scenario \#2, i.e.~%
				private cars with shared parking (with a fixed $r_\textrm{max} = 500\,\mathrm{m}$).}
			\label{pspaces}
		\end{figure*}
		
		\begin{figure}
			\centering
			\includegraphics{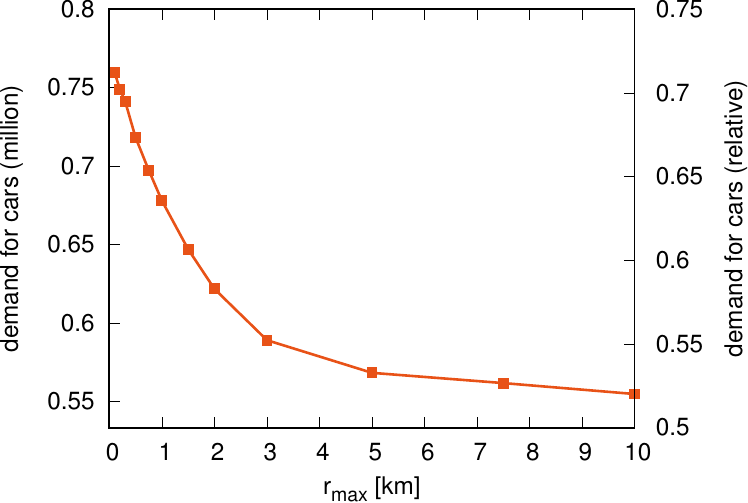}
			\caption{Number of shared cars needed to serve the given mobility demand in scenarios \#3 and \#4, as a function of $r_\mathrm{max}$.
				Absolute numbers are displayed on the left $y$-axis, while relative numbers compared to the case of private cars
				(scenarios~\#1 and~\#2) are displayed on the right $y$-axis.}
			\label{cars}
		\end{figure}
		
		The main result of the presented estimation methodology is the number of cars and city-wide parking spaces needed to cope with the travel
		demand. We display the required number of parking spots in the different scenarios as a function of $r_\textrm{max}$ in Fig.~\ref{pspaces}.	
		We see that for reasonably small values of $r_\textrm{max}$ (i.e.,~between $100\,\mathrm{m}$ and 
		$500\,\mathrm{m}$), around 23\% of parking spaces can be saved by using private cars and sharing parking spaces, as in scenario \#2 (we note 
		that a real city will be between \#1 and \#2, but we expect that most people still have reserved parking). If we introduce shared cars as 
		well (scenario \#3), the reduction in parking demand approaches 40\%. Just comparing the case of private and shared cars (\#2 and \#3), we 
		see that introducing shared cars saves around 20\% of parking spaces from an already highly optimized system with shared parking (see the
		inset in the right panel of Fig.~\ref{pspaces}).
		
		For private or shared cars driven by their users, the $r_\textrm{max}$ distance is essentially the maximum distance people are willing
		to walk from their parking spot to their final destination. In our main simulations, we considered $r_\textrm{max}$ between
		$100\,\mathrm{m}$ to $1.5\,\mathrm{km}$ for scenarios \#2 and \#3 and $r_\textrm{max}$ up to $10\,\mathrm{km}$ for scenario \#4 as shown
		in Fig.~\ref{pspaces}. We believe that actually only the smallest values of $r_\textrm{max}$ are realistic for walking; while previous
		studies for transit usage typically consider $500\,\mathrm{m}$ as an acceptable walking distance~\cite{Guerra2012,ITDP2017}, empirical
		studies on parking usually reveal distance to the destination as the main factor when deciding where to park (more important than price,
		time spent searching for parking, etc.)~\cite{Ma2013,Glasnapp2014,Fiez2017}. For this reason, only the leftmost values in Fig.~\ref{pspaces}
		can be considered significant in these cases. We present larger values mainly for comparison between Algorithms~\ref{estimate_private}
		and~\ref{estimate_shared}.
		
		On the other hand, in the case of self-driving vehicles, much larger $r_\textrm{max}$ values are feasible. The results of our analysis
		show that, already with $r_\textrm{max} = 1.5\,\mathrm{km}$, parking needs reduction is above 50\% compared to scenario~\#1, and around
		33\% when compared to scenario~\#2 with $r_\textrm{max} = 500\,\mathrm{m}$, a realistic upper bound on walking. When considering larger
		values of $r_\mathrm{max}$ up to $10\,\mathrm{km}$, savings in parking demand increase up to 63\% (50\% compared to scenario~\#2). However,
		these savings would come at the expense of increased traffic, as discussed in the next section. We note that actual walking or extra travel
		distances can be smaller than $r_\textrm{max}$, which only specifies the upper bound. In Fig.~S6 in the Supplementary Material, we
		display the distribution of actual walking distances in a few typical cases of $r_\textrm{max}$. For most trips, we find that the actual
		walking distance is much lower than the $r_\textrm{max}$ parameter used for the simulation; on the other hand, a relatively small chance
		of having to walk excess distances could be still highly discouraging for potential users.
		
		The fleet size resulting from our estimations is reported in Fig.~\ref{cars}; we see that we can achieve about 30\% reduction with shared
		cars and small $r_\mathrm{max}$
		values suitable for walking, while these gain increase to over 45\% for larger $r_\mathrm{max}$ values achievable with self-driving.
		
		\subsection{Varying simulation parameters}
		
		So far, we have presented results for a limited set of parameters modeling commuting in Singapore. To estimate the robustness of the
		presented results to changes in the simulation parameters, we repeated the experiments for several different parameter combinations. 
		
\begin{figure}
			\includegraphics{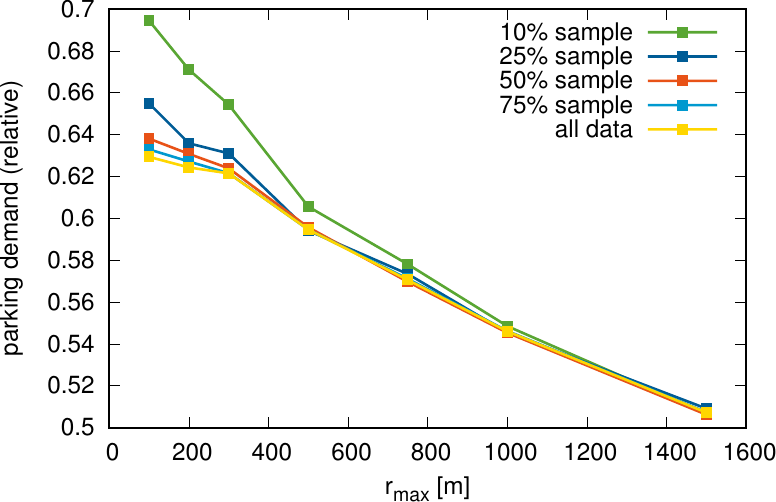}
			\caption{Relative parking demand, compared to scenario $\#1$, for different adoption rates of shared mobility. A sample size (rate) of 10\% results in somewhat less efficiency; above that, we observe gains similar to those obtained with full adoption of shared mobility.}
			\label{pspaces_sampled}
		\end{figure}

		First, we considered different penetration rates of shared mobility, repeating the simulations for scenario \#3/\#4 while varying the number
		of commuters, out of the total number of commuters considered, who use shared vehicles. The results are reported in Fig.~\ref{pspaces_sampled}.
		We see that the possible relative gains (in terms of parking spaces) barely change when at least 25\% (i.e.~about 267,000) of people
		participate in a shared mobility scheme; a smaller sample of only 10\% of people (107,000 people) would instead result in noticeably
		smaller gains (about 5\% difference) when using a radius of $r_\mathrm{max} = 300\, \mathrm{m}$, which we consider a reasonable value
		for walking. On the other hand, for radii of at least $500\,\mathrm{m}$, the gains in parking efficiency are only slightly worse even in
		this case, suggesting that for self-driving cars, a relatively low adoption rate would already bring significant benefits. We note that
		actual gains might be even better as a smaller fleet could be occupied to a larger degree during the day outside commuting hours,
		performing taxi-like service as well.
		
\begin{figure}
			\includegraphics{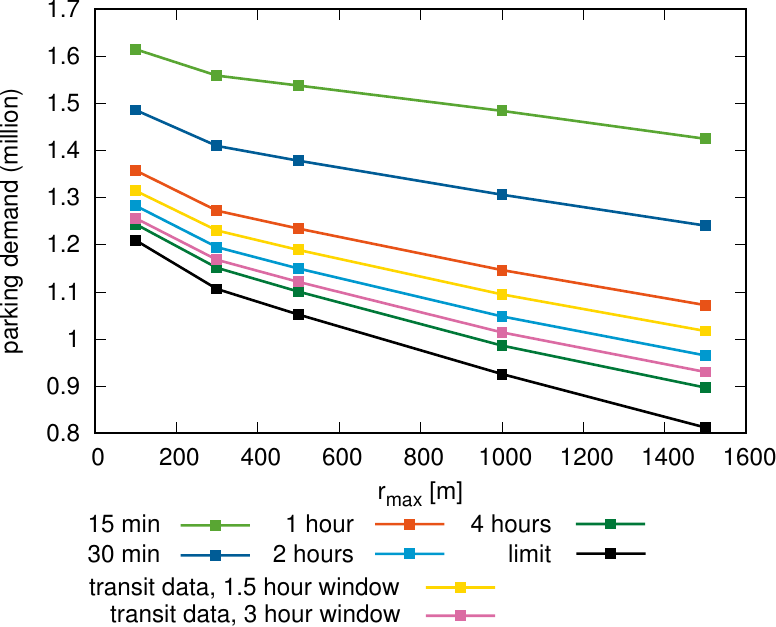}
			\caption{Parking demand for different values of the time window parameter $t_W$. The plot also reports values obtained when trip starting times are randomly generated according to probability distribution extracted from transit data.}
			\label{pspaces_tw}
		\end{figure}

		Furthermore, we repeated the simulations using Algorithm~\ref{estimate_shared} for several different commute lengths of the time window
		$t_W$. These results, reported in Fig.~\ref{pspaces_tw}, indicate that $t_W$ is indeed an important parameter as a commute windows value
		below one hour significantly decreases sharing opportunities. On the other hand, higher values of the commute window will
		only add moderate reductions in parking needs. Furthermore, using travel timings generated from the transit data does not alter the results
		significantly. We note that the one hour commute window used to obtain the main results of this paper can be still considered a
		conservative estimate (e.g.~the activity peaks seen in transit data seem significantly longer as we show in Fig.~S4 in the Supplementary
		Material).
		
		We compute a further measure to characterize the inherent inefficiency due to unbalanced commute flows. This bound is obtained by
		applying the same model under the assumption of instanstaneouos travel (i.e.~all trip times are set to zero, but trips are processed
		in a random order). Since no vehicles are in transit to their destination in this model, the result of this process can be intended as a
		measure of the inherent inefficiency in parking needs that arises from the mere spatial distribution of trip origins and destinations. 
		This is displayed as the black line in Fig.~\ref{pspaces_tw}; we see that there is about 20\% -- 30\% difference between the main results
		(considering $t_W = 1$ hour) and this theoretical limit.

	\subsection{Estimating induced extra miles traveled}
		
		\begin{figure}
			\includegraphics{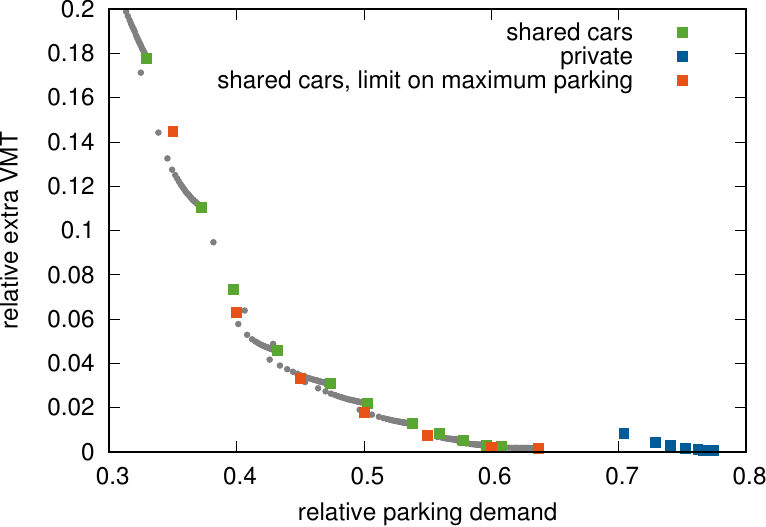}
			\caption{Relative extra traffic, measured as vehicle miles traveled (VMT) as a function of relative reduction in parking needs.
				The blue and green point are the results from the simulation run as Algorithms~1 and~2 for private cars (\#2) and shared or
				self-driving cars (\#3 or \#4) respectively. The points displayed are the results obtained after running the simulation for
				30 days. The grey points are the results for shared or self-driving cars on every individual day up to the main results; the
				number of required parking spaces increases over the course of the simulation, while the extra traffic decreases. The red
				points are results of a modified simulation where the maximum number of available parking spaces is a fixed. For an
				explanation of these methodological differences, refer to the Supplementary Material.}
			\label{pspaces_dist}
		\end{figure}
		
		Self-driving cars would allow parking farther away from the origin or destination of a trip. We have seen that this would further
		reduce the number of parking spaces required (Figs.~\ref{pspaces} and~\ref{cars}). This benefit nevertheless
		comes at a price of increased traffic, which we quantify here as an increase in the total vehicle miles traveled (VMT). In this section,
		we present results for estimating this extra VMT to be able to find a good trade-off between less parking (and cars) and more traffic.
		To obtain this, during the course of the simulation, we recorded the distances between the origin or destination of a trip and the parking
		spot used; we sum these distances and compare them to the total distance that people have to travel between their home and work locations.
		We present the relative extra distance traveled as a function of the previously established reduction in parking demand and also in a
		slightly modified case where the maximum number of parking spots is capped at a number determined from previous simulation runs (see the
		Supplementary Material for more explanation on this). We see that using self-driving vehicles, achieving about 50\% reduction in parking
		space requirement over scenario \#1 will only add about 2\% extra VMT, while further gains come at the cost of potentially significantly
		more vehicle travel. How this extra travel affects congestion will depend on how self-driving cars perform in real-world traffic, i.e.,~%
		whether they can compensate increased traffic by being more efficient.
		
		We note that allowing longer distances (and more traffic) can correspond to a scenario where instead of on-site parking garages,
		operators of self-driving fleets have depots placed in strategic locations in the city. Assuming a fleet of interchangeable vehicles
		(or a few vehicle types), these depots can be highly efficient, have a much smaller total area than traditional parking garages~%
		\cite{Nourinejad2018}. This would present further reductions in the footprint of parking in cities, introducing both opportunities
		and challenges in re-using existing parking facilities.

	\section{Discussion}
		
		In this paper, we evaluated the possible gains in parking demand if a significant number of commuters switched from private
		cars to shared or self-driving vehicles. We focused explicitely on home-work commuting as these trips contribute a large portion
		of traffic, are highly unbalanced, and reserved parking at home and work locations take up huge amount of space in cities. We
		used a large sample of commuters in Singapore, for whom we obtained home and work locations from a mobile phone dataset.
		We evaluated the effect of sharing parking, sharing cars, and using shared self-driving cars on the number of parking spaces required.
		We found that with self-driving cars, about 50\% reduction of parking needs is possible with allowing only 2\% more travel (VMT) due
		to cars traveling to and from parking spaces that now need not be placed on site for all home and work locations. We expect that
		further trips during the day could be served with only minimal extra cars and parking, potentially providing even higher benefits in
		efficiency as currently there could be as many as 3 parking spaces per car in a city.
		
		We note that the main practical factor affecting the reported gains is the shared nature of vehicles. From a technical point
		of view, whether these vehicles are self-driving seems to have effect only on the reasonable values of the main parameter $r_\textrm{max}$
		in our model. On the other hand, there is a large conceptual difference between the two cases, where self-driving vehicles have further
		advantages. Since for conventional cars the $r_\textrm{max}$ parameter represents walking distance, we expect people's expectations to
		be quite low, and also high dissatisfaction if this expectation is ever exceeded. This could also lead to an anxiety, which can deter
		potential users from relying on shared cars as their primary means of transportation. On the other hand, an operator of a fleet of
		self-driving cars has much more flexibility in choosing an $r_\textrm{max}$ value and also can provide better worst-case guarantees on
		vehicle availability. As finding parking is no longer the users' responsibility, this can even present advantages over private vehicles.
		Furthermore, rebalancing is much simplified if no human employees are required. 
		
		Based on these factors, we find it reasonable that the adoption of conventional car-sharing has been relatively slow. On the other hand, 
		we can expect the adoption of shared self-driving cars to take up much faster once the technology is deployed on commercial scales. Thus, 
		we can expect that large areas which are currently dedicated to parking will be freed up in the near future. We note that repurposing 
		existing infrastructure, especially underground parking facilities, can be challenging. On the other hand, repurposing of existing
		parking space can be especially attractive for logistics and light industrial use, which currently cannot afford such central locations. 
		
		Future work is 
		necessary to assess the full impact on traffic congestion and total parking needs due to potentially changing habits and transportation 
		mode choices as a result of the introduction of self-driving cars, which were not modeled in the current work. We finally note that our 
		simulation methodology can be easily adapted to more detailed datasets, e.g.,~logs of individual trips; using these would provide even 
		more accurate predictions on the effect that shared and self-driving cars can have on parking demand.
		
		\section*{Acknowledgment}
			
			The authors thank all sponsors and partners of the MIT Senseable City Laboratory including Allianz, the Amsterdam Institute for 
			Advanced Metropolitan Solutions, the Fraunhofer Institute, Kuwait-MIT Center for Natural Resources and the Environment, 
			Singapore-MIT Alliance for Research and Technology (SMART) and all the members of the Consortium.

\bibliographystyle{unsrtnat}

\begin{thebibliography}{39}
\small
\providecommand{\natexlab}[1]{#1}
\providecommand{\url}[1]{\texttt{#1}}
\providecommand{\doi}[1]{\url{https://doi.org/#1}}

\bibitem[Chester et~al.(2015)Chester, Fraser, Matute, Flower, and
  Pendyala]{Chester2015}
Mikhail Chester, Andrew Fraser, Juan Matute, Carolyn Flower, and Ram Pendyala.
\newblock {Parking Infrastructure: A Constraint on or Opportunity for Urban
  Redevelopment? A Study of Los Angeles County Parking Supply and Growth}.
\newblock \emph{Journal of the American Planning Association}, 81\penalty0
  (4):\penalty0 268--286, 2015.
\newblock ISSN 0194-4363.
\newblock \doi{10.1080/01944363.2015.1092879}.

\bibitem[Flyvbjerg(2007)]{Flyvbjerg2007}
Bent Flyvbjerg.
\newblock {Policy and planning for large-infrastructure projects: Problems,
  causes, cures}.
\newblock \emph{Environment and Planning B: Planning and Design}, 34\penalty0
  (4):\penalty0 578--597, 2007.
\newblock ISSN 02658135.
\newblock \doi{10.1068/b32111}.

\bibitem[Mercier et~al.(2016)Mercier, Carrier, Duarte, and
  Tremblay-Racicot]{Mercier}
Jean Mercier, Mario Carrier, F{\'{a}}bio Duarte, and Fanny Tremblay-Racicot.
\newblock {Policy tools for sustainable transport in three cities of the
  Americas: Seattle, Montreal, and Curitiba}.
\newblock \emph{Transport Policy}, 50:\penalty0 95--105, 2016.
\newblock ISSN 0967-070X.
\newblock \doi{10.1016/j.tranpol.2016.06.005}.

\bibitem[Shoup(2005)]{Shoup}
Donald~C Shoup.
\newblock \emph{{The high cost of free parking}}.
\newblock Planners Press Chicago, 2005.
\newblock ISBN 1884829988.

\bibitem[Weinberger(2012)]{Weinberger2012}
Rachel Weinberger.
\newblock {Death by a thousand curb-cuts: Evidence on the effect of minimum
  parking requirements on the choice to drive}.
\newblock \emph{Transport Policy}, 20:\penalty0 93--102, 2012.
\newblock ISSN 0967070X.
\newblock \doi{10.1016/j.tranpol.2011.08.002}.

\bibitem[McCahill et~al.(2016)McCahill, Garrick, Atkinson-Palombo, and
  Polinski]{McCahill2016}
Christopher~T. McCahill, Norman Garrick, Carol Atkinson-Palombo, and Adam
  Polinski.
\newblock {Effects of Parking Provision on Automobile Use in Cities}.
\newblock \emph{Transportation Research Record: Journal of the Transportation
  Research Board}, 2543:\penalty0 159--165, 2016.
\newblock ISSN 0361-1981.
\newblock \doi{10.3141/2543-19}.

\bibitem[Fiez and Ratliff(2017)]{Fiez2017}
Tanner Fiez and Lillian Ratliff.
\newblock {Data-Driven Spatio-Temporal Analysis of Curbside Parking Demand: A
  Case-Study in Seattle}.
\newblock 2017.
\newblock URL \url{http://arxiv.org/abs/1712.01263}.

\bibitem[Fagnant and Kockelman(2015)]{Fagnant2015}
Daniel~J. Fagnant and Kara Kockelman.
\newblock {Preparing a nation for autonomous vehicles: Opportunities, barriers
  and policy recommendations}.
\newblock \emph{Transportation Research Part A: Policy and Practice},
  77:\penalty0 167--181, 2015.
\newblock ISSN 09658564.
\newblock \doi{10.1016/j.tra.2015.04.003}.

\bibitem[Krueger et~al.(2016)Krueger, Rashidi, and Rose]{Krueger2016}
Rico Krueger, Taha~H. Rashidi, and John~M. Rose.
\newblock {Preferences for shared autonomous vehicles}.
\newblock \emph{Transportation Research Part C: Emerging Technologies},
  69:\penalty0 343--355, 2016.
\newblock ISSN 0968090X.
\newblock \doi{10.1016/j.trc.2016.06.015}.

\bibitem[Harper et~al.(2016)Harper, Hendrickson, Mangones, and
  Samaras]{Harper2016}
Corey~D. Harper, Chris~T. Hendrickson, Sonia Mangones, and Constantine Samaras.
\newblock {Estimating potential increases in travel with autonomous vehicles
  for the non-driving, elderly and people with travel-restrictive medical
  conditions}.
\newblock \emph{Transportation Research Part C: Emerging Technologies},
  72:\penalty0 1--9, 2016.
\newblock ISSN 0968090X.
\newblock \doi{10.1016/j.trc.2016.09.003}.

\bibitem[Daziano et~al.(2017)Daziano, Sarrias, and Leard]{Daziano2017}
Ricardo~A. Daziano, Mauricio Sarrias, and Benjamin Leard.
\newblock {Are consumers willing to pay to let cars drive for them? Analyzing
  response to autonomous vehicles}.
\newblock \emph{Transportation Research Part C: Emerging Technologies},
  78:\penalty0 150--164, 2017.
\newblock ISSN 0968090X.
\newblock \doi{10.1016/j.trc.2017.03.003}.

\bibitem[van~den Berg and Verhoef(2016)]{VandenBerg2016}
Vincent~A.C. van~den Berg and Erik~T. Verhoef.
\newblock {Autonomous cars and dynamic bottleneck congestion: The effects on
  capacity, value of time and preference heterogeneity}.
\newblock \emph{Transportation Research Part B: Methodological}, 94:\penalty0
  43--60, 2016.
\newblock ISSN 01912615.
\newblock \doi{10.1016/j.trb.2016.08.018}.

\bibitem[Handy(2005)]{Handy2005}
Susan Handy.
\newblock {Smart Growth and the Transportation-Land Use Connection: What Does
  the Research Tell Us?}
\newblock \emph{International Regional Science Review}, 28\penalty0
  (2):\penalty0 146--167, 2005.
\newblock ISSN 0160-0176.
\newblock \doi{10.1177/0160017604273626}.

\bibitem[Gupta et~al.(2006)Gupta, Kalmanje, and Kockelman]{Gupta2006}
Surabhi Gupta, Sukumar Kalmanje, and Kara~M. Kockelman.
\newblock {Road pricing simulations: Traffic, land use and welfare impacts for
  Austin, Texas}.
\newblock \emph{Transportation Planning and Technology}, 29\penalty0
  (1):\penalty0 1--23, 2006.
\newblock ISSN 03081060.
\newblock \doi{10.1080/03081060600584130}.

\bibitem[Smith(2012)]{Smith2012}
Bryant~Walker Smith.
\newblock {Managing Autonomous Transportation Demand}.
\newblock \emph{Santa Clara Law Review}, 52\penalty0 (4):\penalty0 1401--1422,
  2012.
\newblock ISSN 02729490.
\newblock \doi{10.1525/sp.2007.54.1.23.}

\bibitem[Firnkorn and M{\"{u}}ller(2012)]{Firnkorn2012}
J{\"{o}}rg Firnkorn and Martin M{\"{u}}ller.
\newblock {Selling Mobility instead of Cars: New Business Strategies of
  Automakers and the Impact on Private Vehicle Holding}.
\newblock \emph{Business Strategy and the Environment}, 21\penalty0
  (4):\penalty0 264--280, 2012.
\newblock ISSN 09644733.
\newblock \doi{10.1002/bse.738}.

\bibitem[Fagnant and Kockelman(2014)]{Fagnant2014}
Daniel~J. Fagnant and Kara~M. Kockelman.
\newblock {The travel and environmental implications of shared autonomous
  vehicles, using agent-based model scenarios}.
\newblock \emph{Transportation Research Part C: Emerging Technologies},
  40:\penalty0 1--13, 2014.
\newblock ISSN 0968090X.
\newblock \doi{10.1016/j.trc.2013.12.001}.

\bibitem[Burns et~al.(2013)Burns, Jordan, and Scarborough]{Burns2013}
Lawrence~D. Burns, William~C. Jordan, and Bonnie~a. Scarborough.
\newblock {Transforming Personal Mobility}.
\newblock Technical report, The Earth Institute, Columbia University, 2013.
\newblock URL
  \url{http://wordpress.ei.columbia.edu/mobility/files/2012/12/Transforming-Personal-Mobility-Aug-10-2012.pdf}.

\bibitem[Brownell and Kornhauser(2014)]{Brownell2014}
Chris Brownell and Alain Kornhauser.
\newblock {A Driverless Alternative Fleet Size and Cost Requirements for a
  Statewide Autonomous Taxi Network in New Jersey}.
\newblock \emph{Transportation Research Record}, 2416:\penalty0 73--81, 2014.
\newblock \doi{10.3141/2416-09}.

\bibitem[Jorge and Correia(2013)]{Jorge2013}
Diana Jorge and Gon{\c{c}}alo Correia.
\newblock {Carsharing systems demand estimation and defined operations: A
  literature review}.
\newblock \emph{European Journal of Transport and Infrastructure Research},
  13\penalty0 (3):\penalty0 201--220, 2013.
\newblock ISSN 15677141.
\url{http://www.ejtir.tudelft.nl/issues/2013_03/pdf/2013_03_02.pdf}

\bibitem[Kek et~al.(2006)Kek, Cheu, and Chor]{Kek2006}
Alvina Kek, Ruey Cheu, and Miaw Chor.
\newblock {Relocation Simulation Model for Multiple-Station Shared-Use Vehicle
  Systems}.
\newblock \emph{Transportation Research Record}, 1986\penalty0 (1):\penalty0
  81--88, 2006.
\newblock ISSN 0361-1981.
\newblock \doi{10.3141/1986-13}.

\bibitem[Papanikolaou(2011)]{Papanikolaou2011}
Dimitris Papanikolaou.
\newblock {A new system dynamics framework for modeling behavior of vehicle
  sharing systems}.
\newblock \emph{Proceedings of the 2011 Symposium on Simulation for
  Architecture and Urban Design, Society for Computer Simulation International,
  Boston, Massachusetts}, pages 126--133, 2011.
\newblock URL \url{http://dl.acm.org/citation.cfm?id=2048552}.

\bibitem[Anderson(2000)]{Anderson2000}
J~E Anderson.
\newblock {A review of the state of the art of personal rapid transit}.
\newblock \emph{Journal of advanced transportation}, 34\penalty0 (1):\penalty0
  3--29, 2000.
\newblock ISSN 2042-3195.
\newblock \doi{10.1002/atr.5670340103}.

\bibitem[Martin et~al.(2010)Martin, Shaheen, and Lidicker]{Martin2010}
Elliot Martin, Susan Shaheen, and Jeffrey Lidicker.
\newblock {Impact of Carsharing on Household Vehicle Holdings}.
\newblock \emph{Transportation Research Record: Journal of the Transportation
  Research Board}, 2143:\penalty0 150--158, 2010.
\newblock ISSN 0361-1981.
\newblock \doi{10.3141/2143-19}.

\bibitem[Ballantyne et~al.(2014)Ballantyne, Zhang, Morton, Pavone, Case,
  Spieser, Treleaven, and Frazzoli]{Ballantyne2014}
Kyle Ballantyne, Rick Zhang, Daniel Morton, Marco Pavone, Systems~a Case, Kevin
  Spieser, Kyle Treleaven, and Emilio Frazzoli.
\newblock {Toward a Systematic Approach to the Design and Evaluation of
  Automated Mobility-on-Demand Systems: A Case Study in Singapore}.
\newblock In \emph{Road Vehicle Automation}, pages 229--245. 2014.
\newblock \doi{10.1007/978-3-319-05990-7_20}.

\bibitem[Marczuk et~al.(2015)Marczuk, Hong, Azevedo, Adnan, Pendleton,
  Frazzoli, and Lee]{Marczuk2015}
Katarzyna~Anna Marczuk, Harold Soh~Soon Hong, Carlos Miguel~Lima Azevedo,
  Muhammad Adnan, Scott~Drew Pendleton, Emilio Frazzoli, and Der~Horng Lee.
\newblock {Autonomous mobility on demand in SimMobility: Case study of the
  central business district in Singapore}.
\newblock \emph{Proceedings of the 2015 7th IEEE International Conference on
  Cybernetics and Intelligent Systems, CIS 2015 and Robotics, Automation and
  Mechatronics, RAM 2015}, pages 167--172, 2015.
\newblock \doi{10.1109/ICCIS.2015.7274567}.

\bibitem[Nourinejad et~al.(2018)Nourinejad, Bahrami, and
  Roorda]{Nourinejad2018}
Mehdi Nourinejad, Sina Bahrami, and Matthew~J. Roorda.
\newblock {Designing parking facilities for autonomous vehicles}.
\newblock \emph{Transportation Research Part B: Methodological}, 109:\penalty0
  110--127, 2018.
\newblock ISSN 01912615.
\newblock \doi{10.1016/j.trb.2017.12.017}.

\bibitem[Zhang et~al.(2015)Zhang, Guhathakurta, Fang, and Zhang]{Zhang2015}
Wenwen Zhang, Subhrajit Guhathakurta, Jinqi Fang, and Ge~Zhang.
\newblock {Exploring the impact of shared autonomous vehicles on urban parking
  demand: An agent-based simulation approach}.
\newblock \emph{Sustainable Cities and Society}, 19:\penalty0 34--45, 2015.
\newblock ISSN 22106707.
\newblock \doi{10.1016/j.scs.2015.07.006}.

\bibitem[Zhang and Guhathakurta(2017)]{Zhang2017}
Wenwen Zhang and Subhrajit Guhathakurta.
\newblock {Parking Spaces in the Age of Shared Autonomous Vehicles: How Much
  Parking Will We Need and Where?}
\newblock \emph{Transportation Research Board 96th Annual Meeting}, pages
  17--05399, 2017.
\newblock URL \url{https://trid.trb.org/view.aspx?id=1439127}.

\bibitem[{OECD International Transport Forum}(2015)]{OECD2015}
{OECD International Transport Forum}.
\newblock {Urban Mobility System Upgrade: How shared self-driving cars could
  change city traffic}.
\newblock Technical report, 2015.
\newblock URL
  \url{http://www.internationaltransportforum.org/Pub/pdf/15CPB\_Self-drivingcars.pdf}.

\bibitem[Bagrow and Lin(2012)]{Bagrow2012}
James~P. Bagrow and Yu-Ru Lin.
\newblock {Mesoscopic Structure and Social Aspects of Human Mobility}.
\newblock \emph{PLoS ONE}, 7\penalty0 (5):\penalty0 e37676, 2012.
\newblock ISSN 1932-6203.
\newblock \doi{10.1371/journal.pone.0037676}.

\bibitem[Jiang et~al.(2015)Jiang, Ferreira, and Gonz{\'{a}}lez]{Jiang}
Shan Jiang, Joseph Ferreira, and Marta~C Gonz{\'{a}}lez.
\newblock {Activity-Based Human Mobility Patterns Inferred from Mobile Phone
  Data: A Case Study of Singapore}.
\newblock \emph{IEEE Transactions on Big Data}, 3 (2): 208--219, 2017.
\newblock \doi{10.1109/TBDATA.2016.2631141}

\bibitem[Alexander et~al.(2015)Alexander, Jiang, Murga, and
  Gonz{\'{a}}lez]{Alexander2015}
Lauren Alexander, Shan Jiang, Mikel Murga, and Marta~C. Gonz{\'{a}}lez.
\newblock {Origin-destination trips by purpose and time of day inferred from
  mobile phone data}.
\newblock \emph{Transportation Research Part C: Emerging Technologies},
  58:\penalty0 240--250, 2015.
\newblock ISSN 0968090X.
\newblock \doi{10.1016/j.trc.2015.02.018}.

\bibitem[Shoup(2006)]{Shoup2006}
Donald~C. Shoup.
\newblock {Cruising for parking}.
\newblock \emph{Transport Policy}, 13\penalty0 (6):\penalty0 479--486, 2006.
\newblock ISSN 0967070X.
\newblock \doi{10.1016/j.tranpol.2006.05.005}.

\bibitem[Dowling et~al.(2017)Dowling, Fiez, Ratliff, and Zhang]{Dowling2017}
Chase Dowling, Tanner Fiez, Lillian Ratliff, and Baosen Zhang.
\newblock {How Much Urban Traffic is Searching for Parking?}
\newblock \emph{arXiv preprint}, arXiv:1702:06156, 2017.
\newblock URL \url{http://arxiv.org/abs/1702.06156}.

\bibitem[Guerra et~al.(2012)Guerra, Cervero, and Tischler]{Guerra2012}
Erick Guerra, Robert Cervero, and Daniel Tischler.
\newblock {Half-Mile Circle; Does It Best Represent Transit Station
  Catchments?}
\newblock \emph{Transportation Research Record: Journal of the Transportation
  Research Board}, 2276\penalty0 (2276):\penalty0 101--109, 2012.
\newblock ISSN 0361-1981.
\newblock \doi{10.3141/2276-12}.

\bibitem[ITDP(2017)]{ITDP2017}
ITDP.
\newblock {TOD Standard, v3}.
\newblock Technical report, 2017.
\newblock URL \url{https://www.itdp.org/tod-standard/}.

\bibitem[Ma et~al.(2013)Ma, Sun, He, and Chen]{Ma2013}
Xiaolong Ma, Xiaoduan Sun, Yulong He, and Yixin Chen.
\newblock {Parking Choice Behavior Investigation: A Case Study at Beijing Lama
  Temple}.
\newblock \emph{Procedia - Social and Behavioral Sciences}, 96\penalty0
  (Cictp):\penalty0 2635--2642, 2013.
\newblock ISSN 18770428.
\newblock \doi{10.1016/j.sbspro.2013.08.294}.

\bibitem[Glasnapp et~al.(2014)Glasnapp, Du, Dance, Clinchant, Pudlin, Mitchell,
  and Zoeter]{Glasnapp2014}
James Glasnapp, Honglu Du, Christopher Dance, Stephane Clinchant, Alex Pudlin,
  Daniel Mitchell, and Onno Zoeter.
\newblock {Understanding dynamic pricing for parking in Los Angeles: Survey and
  ethnographic results}.
\newblock \emph{Lecture Notes in Computer Science (including subseries Lecture
  Notes in Artificial Intelligence and Lecture Notes in Bioinformatics)}, 8527
  LNCS:\penalty0 316--327, 2014.
\newblock ISSN 16113349.
\newblock \doi{10.1007/978-3-319-07293-7_31}.

\end{thebibliography}



\section*{Supplementary material}

# Estimating savings in parking demand using shared vehicles for home-work commuting

## Supplementary Material

Dániel Kondor, Hongmou Zhang, Remi Tachet, Paolo Santi, Carlo Ratti

Senseable City Laboratory, MIT, Cambridge MA 02139 USA

June 18, 2018

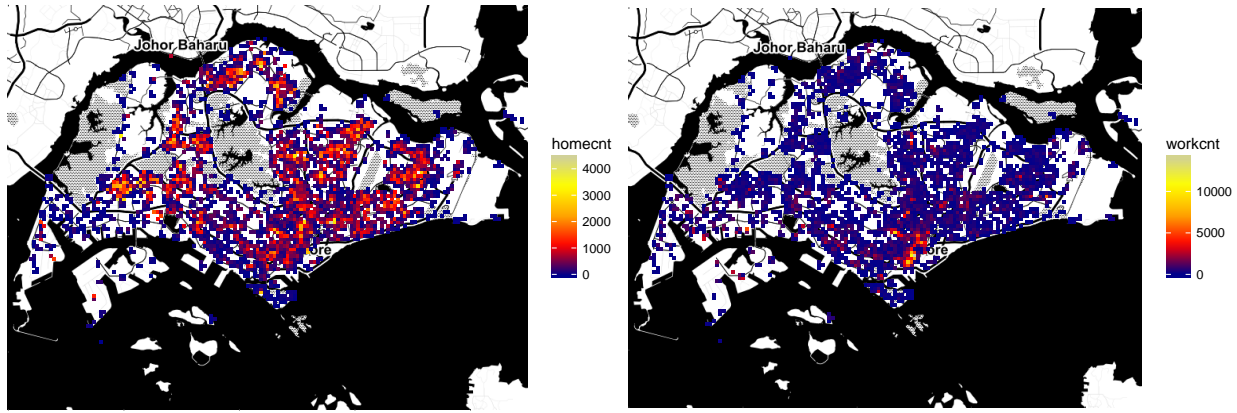

Figure S1: Spatial distribution of home (left) and work (right) locations in Singapore. Note that the distribution of home locations was verified using census data available in Singapore.

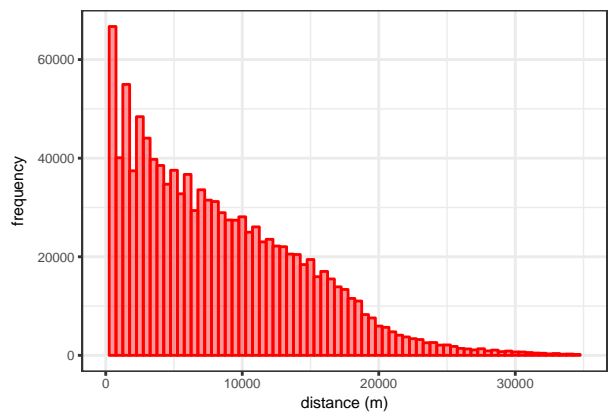

Figure S2: Distribution of travel distances between home and work locations (calculated as simple Euclidean distances). The mean distance is 8.4 km, the median distance is 7.2 km.

## Commute timings

*Transit-based commute times model:* For every user in the dataset, a random time for going to work is selected from the distribution of transit usage data displayed in Fig. S4. Similarly, a random time is selected for going home after work. These random times are constrained between a realistic "commute window" in this case as well, which we select to include the distinctive main peaks of commuting behavior in the data: these are between 6am and 9am for the morning commute and between 5pm and 8pm for the afternoon commute. These are represented by the shaded areas in Fig. S4.

A realistic estimate on when people travel to work or home can have a significant impact on the shareability. While there was a lot of previous work on estimating commutes from mobile phone data as well, we now use a separate, one

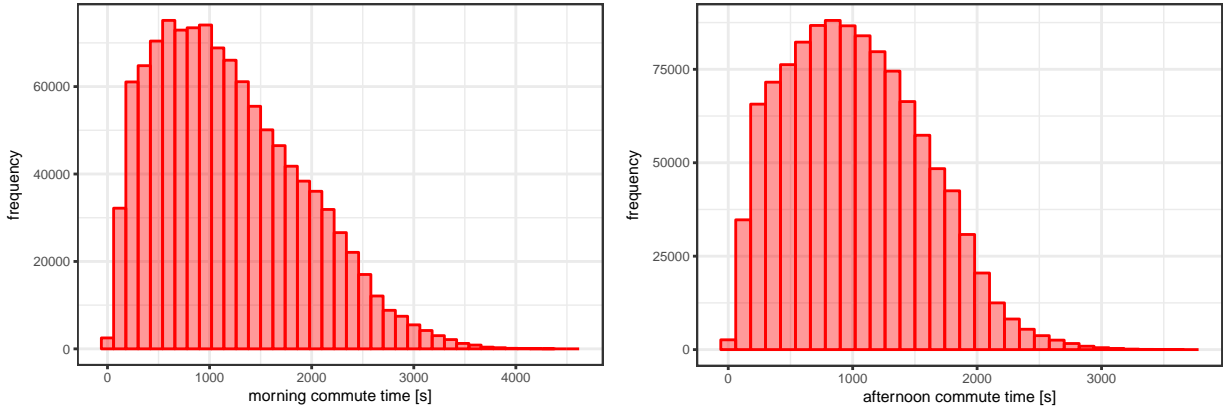

Figure S3: **Distribution of travel times for commuters.** On the left, the distribution of travel times from work to home is displayed; on the right, the distribution of travel times from work to home. The mean of the distributions is 1199 s and 1027 s respectively, while the median is 1090 s and 983 s.

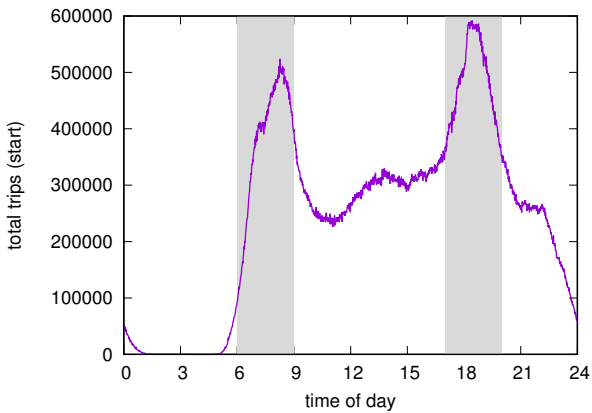

Figure S4: **Temporal distribution of transit usage.** Transit trip start times on an average weekday are displayed. The shaded regions are used in estimating commute start times when validating some of our findings later.

week dataset of public transportation records supplied by Singapore's Land Transportation Authority (LTA) to estimate the typical times of day people travel to work and back home. While having the obvious drawback that it requires the assumption that the habits of people using transit and commuting by cars are similar, this method presents an easier and possibly more reliable picture than trying to find individual trips in the mobile phone data. On the other hand, using average commuting times also neglects any possible systematic variation in commute times which might further affect the results. The distribution of trip start times is displayed in Fig. S4.

## Varying simulation length, imposing a strict maximum limit on parking

So far, the results we presented were obtained after running the simulation for 30 days in a row to account for daily stochastic differences in commute patterns. Now, in Fig. S5, we present a case when we run the simulation for longer time intervals. We present results for the cumulative number of parking spots required as a function of time the simulation is run. We see that there is a small but steady increase in the shared or self-driving scenario (#3 and #4), showing resulting configurations after a day are typically inadequate for satisfying the mobility demand on the next day. This raises the question about what is a realistic number of vehicles that we can expect to cope with the long-term mobility demand. We can consider two possible ways to solve the problem of apparent increasing demand in parking as simulation time progresses. One is the obvious possibility of implementing some rebalancing; while this can present a significant cost for operators of car-sharing systems (i.e. scenario #3), in the case of self-driving cars (scenario #4), rebalancing will require only minimal costs; we emphasize that all our results were obtained without including rebalancing.

On the other hand, in the case of self-driving, we can simply further relax the strict requirement that cars should not travel more than $r_{max}$ without a passenger. We note that the $r_{max}$ limit is actually a technical part in our simulation which allows us to evaluate how many parking spots to "add" to the city. In a realistic scenario however, any request by a user would be serviced by the closest car, regardless of the actual distance. While in the case of car-sharing, not having a

2

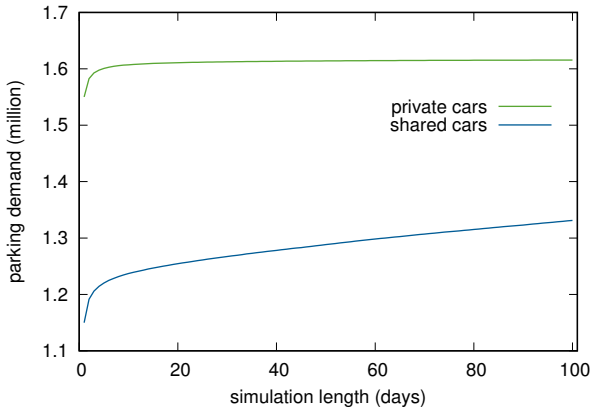

Figure S5: Parking spaces needed to meet the mobility demand as a function of the number of days we run the simulation. We see that the number of parking spaces quickly saturates in the case of private cars, meaning that we can easily account for the stochastic nature of our simulation. In the case of shared vehicles, the number of parking spaces keeps growing, albeit at a slow rate; this implies that we do not reach a stable configuration of cars and in a real system some rebalancing might be necessary.

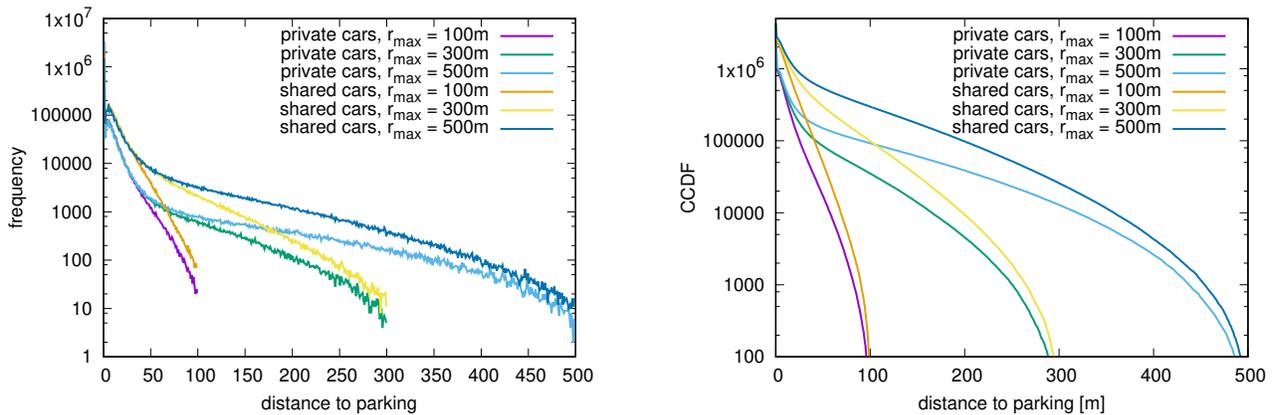

Figure S6: Distribution of distances between trip start and end locations and parking locations. For conventional vehicles, this corresponds to the distance users need to walk, while for AVs, this is the extra distance the vehicle has to travel without a passenger. The left panel displays individual frequencies (binned in 1 m intervals), while the right panel displays the complementer cumulative distribution, i.e. the number of cases where a passenger has to walk *at least* the distance shown on the $x$-axis. Note that the $y$-axis is logarithmic in both panels. The large majority of trips only requires very short extra distances, while there is a small share requiring walking / extra travel up to the limit $r_{\max}$.

car or parking spot available under $r_{\max}$ will result in a user having to walk an excess distance and thus will lead to a high level of user dissatisfaction, in the case of self-driving cars, having a car further than $r_{\max}$ will only result in a slightly increased waiting time for the user in question, which will have a much less effect on user satisfaction with the service (given that it happens rarely). With this in mind, we also implemented a modified version of our simulation algorithm (Algorithm 2 in the main text). In this case, we limit the maximum number of parking spaces to a predetermined amount. After this limit has been reached, the closest car or parking space is selected regardless of the distance to the destination and thus no new parking is added to the system. In this case, the $r_{\max}$ parameter is not a direct determinant of the functioning of the system but a parameter which affects the process of how we distribute the available parking spaces in the city. To accommodate this change, we run the simulation with different $r_{\max}$ values for each limit value we select for parking spaces and select the $r_{\max}$ which minimizes the extra distance traveled in practice. As seen in Fig. 6 in the main text, results obtained this way show a good agreement with the results of the original simulation methodology. This allows us to accept the results presented in Fig. 6 as a good approximation for the trade-off between reduced parking and extra traffic.

## Walking distances

While the $r_{\max}$ distance acts as an upper bound for the distance passengers have to walk (in scenarios #2 and #3) or AVs have to travel without a passenger before the start or after the end of the trip, actual distances will be typically lower as we always select the closest available vehicles or parking. In Fig. S6, we present the distribution of actual distances of the vehicles used to the trip origin and parking spot to trip destination for several different values of $r_{\max}$ in the case of private vehicles (scenario #2) and shared vehicles (either scenario #3 or #4). We see that a large share of trips involves significantly less walking / extra travel than the $r_{\max}$ upper bound. Nevertheless, a non-negligible amount of trips has extra travel close to $r_{\max}$. In the case of conventional vehicles (either scenario #2 or #3) where this is a walking distance, this makes it reasonable to limit $r_{\max}$ to small values.

## Statistical variation in results

All simulations in the current work involve randomness as the timing of trips are randomly generated. To characterize the extent this affects the results, we repeated all simulation runs 100 times and calculated relevant statistics on the results. Generally we find that variations can be well characterized by a normal distribution and the standard deviations are very small compared to the mean values (less than 1% in all cases). We further carried out a Shapiro-Wilk test for normally distributed values and found that a null hypothesis for normal distribution cannot be rejected in almost all cases; the number of cases having low $p$-values is consistent with the large number of total parameter tests, meaning that these could occur randomly. This way, we conclude that an assumption of normally distributed results is reasonable. In Table S1 (available as separate download), we display results for all parameter combinations tested together with standard deviations, Shapiro-Wilk test statistics and $p$-values. In Fig. S7 we display the distribution of result values for the case of shared cars, $t_W = 1\,\mathrm{hour}$ and $r_{\max} = 500\,\mathrm{m}$ as an example of typical cases. Furthermore, to visually assess difference from a normal distribution in some of the worst cases, in Fig. S8, we display the distributions with the two lowest $p$-values. We see that the difference is not systematic and there are no significant outliers in these cases as well. This way, we believe that using the mean and the standard deviation to characterize the results is reasonable. In Fig. S9, we display the cumulative distribution of the $p$-values obtained in the different test cases. We see that these are relatively evenly distributed in the $[0,1]$ interval, supporting that these are obtained with a random process. We note that in total we had results for 196 different parameter combinations for $N_C$, in 316 parameter combinations for $N_P$ and in 321 different parameter combinations for $d_{\mathrm{tot}}$. This is the number of Shapiro-Wilk tests carried out and the number of different $p$-values obtained as well. In this case, finding the low $p$-values can be attributed to randomness as well, thus we believe our choice of modeling the results with normal distributions is reasonable.

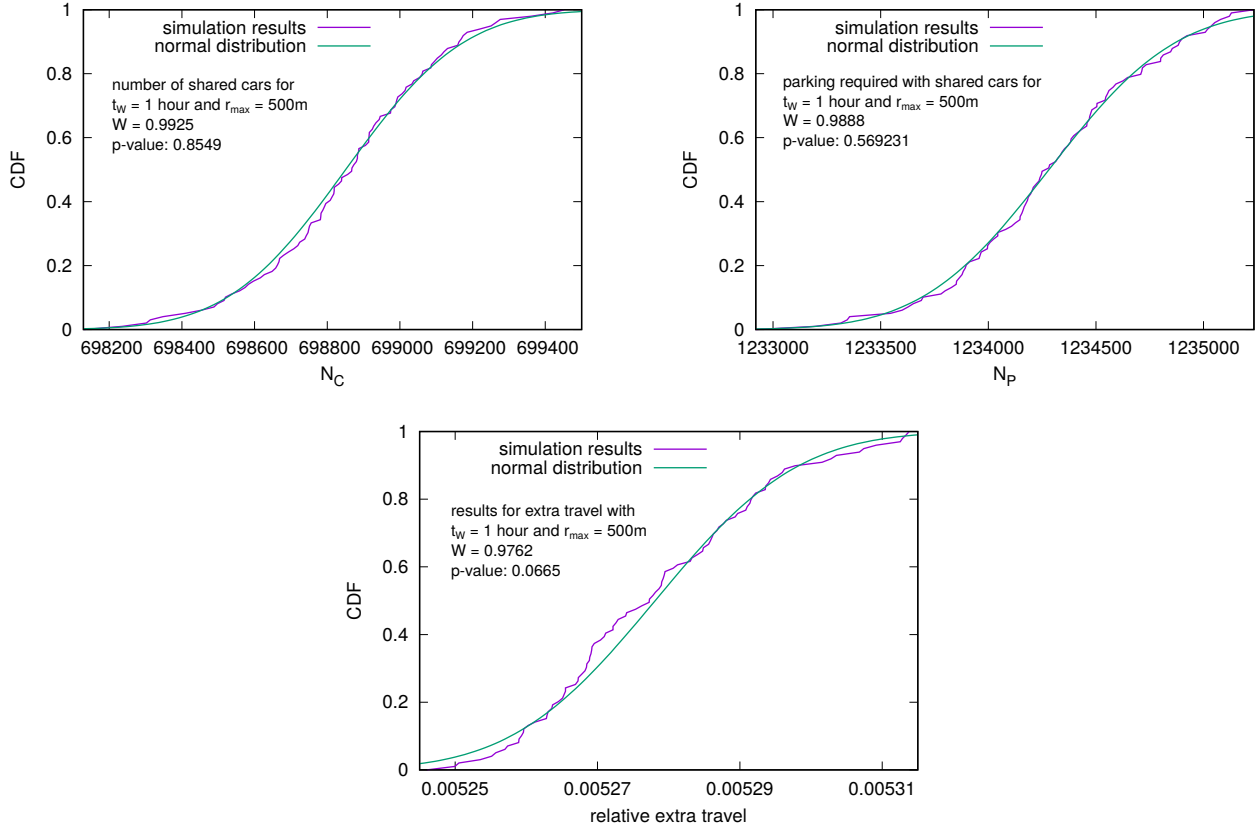

Figure S7: Distribution of results obtained from individual runs with different random trip start times. Example empirical cumulative distribution for shared vehicles (scenario #3 / #4) and parameters $t_W = 1$ hour and $r_{max} = 500$ m is displayed for the number of cars ($N_C$, top left panel), number of parking spaces ($N_P$, top right panel) and extra travel distance (bottom panel). In all cases, all 100 simulation result values in the empirical distribution fall close to the mean (standard deviations are $< 0.04\%$ for the number of cars and parking and $\sim 0.3\%$ for the extra travel distance). The empirical distribution is well approximated with a normal distribution in all cases. The respective normal distributions (i.e. distributions with the same mean and standard deviation as the empirical ones) are shown for comparison as well.

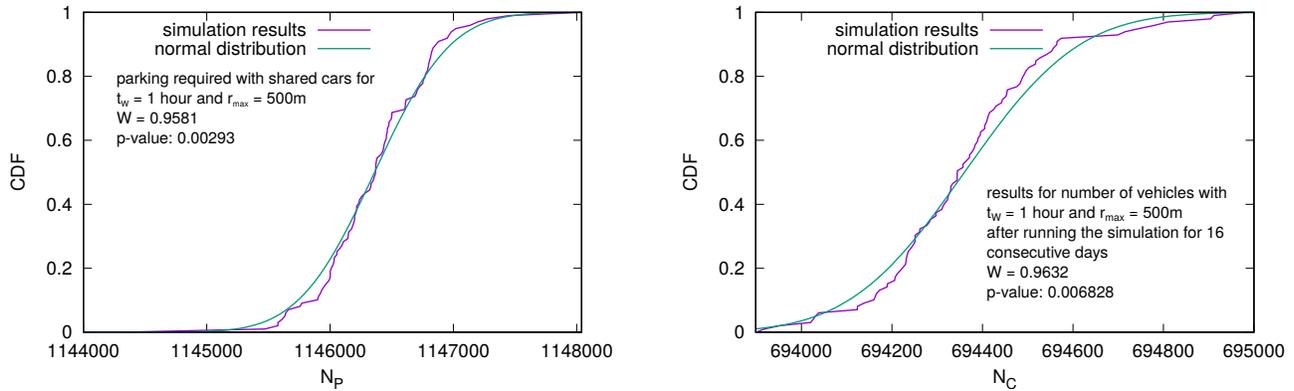

Figure S8: Illustration of the distribution of result values for the two "worst" cases, i.e. where the resulting $p$-values would indicate rejecting a normal distribution. Note that distributions are still relatively close to a normal distribution and the differences are mainly in the shape, without significant outliers. This way, using the mean value as an estimate of result is still justified.

5

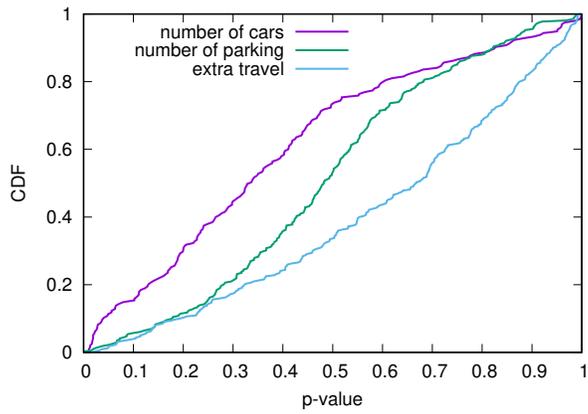

Figure S9: Distribution of *p*-values resulting from all Shapiro-Wilk tests carried out for all different combinations of parameter values. We note that these are quite evenly distributed, thus the lower *p*-values are likely the result of random variation as well.



\end{document}